# Assessing Water Performance Indicators for Leakage Reduction and Asset Management in Water Supply Systems

## G. Mazzolani[1], F.G. Ciliberti[2], L. Berardi[2] and O. Giustolisi[3]

[1]Acquedotto Pugliese s.p.a., Via Cognetti n. 36, 70121 Bari, Italy.
[2]Department of Engineering and Geology, University "G. D'Annunzio" of Chieti Pescara, Pescara, Italy.
[3]Department of Civil, Environmental, Land, Building Engineering and Chemistry, Technical University of Bari, Bari, Italy.

Corresponding author: Orazio Giustolisi (orazio.giustolisi@poliba.it)

**Key Points:**

- This study analyzes two performance indicators that are largely used to drive investments on leakage management and verify the relevant achievements at both system and District Metering Areas (DMAs) scale, namely the linear leakage index and the percentage leakage index.

- It is demonstrated that the percentage leakage index is inappropriate and misleading because of (i) its mathematical structure and (ii) the ambiguity of the definition of water consumption, when looking at the actual utilizations within a supplying system layout.

- A case study from a real system, including supply pipelines, water distribution networks and relevant DMAs is used to demonstrate the inconsistencies of the percentage leakage index, as well as the pertinent use of the linear leakage index.

## Abstract

Water Supply Systems are essential infrastructures for the socio-economic life of urban cities. To improve their reliability, water utilities undertake several short- and long-term operational tasks based on technical and economic constraints. These activities are motivated by many factors, including increasing leakage rates due to infrastructure aging, increased consumer demands and need for sustainable use of water and energy. European and national regulatory bodies have promoted investment programs for allowing water utilities to reach common standards of reliability and quality of service among countries. Targets of management and operational achievements are usually measured using specific performance indicators. The Italian Regulatory Authority for Energy, Networks and Environment (ARERA) recently introduced the Regulation of the technical performances of water utilities. Performances on leakage management and investment plans of the utilities are thus based on two indicators named $M1_a$ (linear leakage index) and $M1_b$ (percentage leakage index). This paper analyzes in details the inconsistencies of the percentage leakage index ($M1_b$), mainly due to its mathematical formulation and the ambiguity of defining water consumption as part of the total system inflow. The discussion is supported by a real case study, where both indicators have been calculated to assess their impact on management decisions and investment plans. The inconsistencies of the percentage leakage index are further demonstrated for various layouts of water supply systems.

## 1 Introduction

Water Supply Systems (WSS) are critical components of modern urban infrastructures, providing essential services to billions of people worldwide. WSS are meant herein as composed of Water Distribution Networks (WDN), where most consumers are connected, and transmission systems, carrying water from waters sources to reservoirs, tanks or pumps, or between neighbouring WDNs, with few or no connection to consumers.

The aging of these infrastructures (Snider & McBean, 2020), the associated increase of water losses and water quality degradation (Frauendorfer & Liemberg, 2010, Alegre et al., 2016) have become the major challenge of water utilities. In Europe, most WDNs are severely affected by water losses (Lallana and Thyssen, 2008), including Spain (González-Gómez et al., 2012), Greece (Karathanasi & Papageorgakopoulos, 2016) and Italy in particular, where water losses reach 42.2% of the total water volume supplied at national scale, with the worst performances in Southern Italy (ISTAT, 2023).

The deterioration of WDNs is a natural process that occurs over time, resulting from the combined effects of aging, corrosion, and other forms of wear and tear. As pipes, connections, devices and other infrastructure components reach their technical service life, they become increasingly susceptible to failure, causing water leakage and other service disruptions (Mays, 2000, Allen et al., 2018). Physical leakages are recognized to decrease the hydraulic capacity of WDNs (van Zyl & Clayton, 2007) and it was proved that larger leakage rate is related to increased rate of pipe breaks (Girard & Stewart, 2007), thus exposing the infrastructure to significant technical impacts, including mismatching of minimum pressure requirements, poor pumping performance and threatening water quality, as well as economical impacts, related to lost revenues from water sales, increased energy costs, along with those for repair and maintenance. Therefore, water utilities need, on one hand, to develop proactive strategies to optimize and monitor the hydraulic behavior of their networks (St. Clair and Sinha, 2011) and, on the other hand, to prioritize plans for leakage reduction and pipes rehabilitation (Mutikanga et al., 2013).

The effectiveness in managing WSS and delivering safe, reliable, and affordable water services to the customers is measured by Water Performance Indicators (WPIs) (Alegre et al., 2016, NWC, 2012). The concept behind WPIs is to incorporate all relevant aspects of WSS management, including physical, economic, operation and service quality factors, and make a cross-comparison with benchmarks of similar water utilities. Such indicators are developed for being easy to calculate and providing the most informative content on the state of networks, depending on the availability of data for their calculations (Lambert et al., 1999). WPIs related to water losses are commonly adopted for assessing the impacts of planning and maintenance activities (Cabrera et al., 2007; Liemberger et al., 2007), as well as to drive the allocation of investments.

The term "water losses" refers, as stated by AWWA (AWWA, 2007), to the sum of "real" losses, i.e. the water volume physically lost through breaks and bursts of transmission and distribution pipes and overflows from tanks, and "apparent" losses, which accounts for metering or billing inaccurancies and unauthorized consumptions. "Apparent" losses are considered a relevant issue in developing countries (Mutikanga et al., 2011), as opposed to the developed countries, where are usually estimated as a small fraction of the inflow volume (AWWA, 2006), due to the lack of standardized methodologies for their assessment. Therefore, without impairing its generality, this paper refers only to the "real" losses (or "leakages") for calculating the indicators and the further hydraulic analysis.

The most frequently adopted WPIs for real water losses refer to the volume of leakages established in the annual WSS water balance, and are related to some physical characteristics of the infrastructure, including length of mains, number of properties and service connections (Alegre et al., 2016).

Among such WPIs, the Infrastructure Leakage Index (ILI) (Lambert, 1999; Lambert & McKenzie, 2002), which has been recommended by IWA, takes into account as well the pressure level in the WDN. Several shortcomings are associated to ILI, including the methodology applied for identifying the WDS pressure (Liemberger, 2002; Seago et al., 2005), the uncertainty of the empirical estimate of UARL (Unavoidable Annual Real Losses) volume, which is independent of pipeline age, materials and diameters, the uncertainties in the confidence levels under different pressure scenarios (Alegre et al., 2016) and the major inconsistencies with respect to actual leakage reduction achievements (Berardi et al., 2018).

WPIs have been adopted from regulatory bodies (Alegre et al., 2016; Carpenter et al., 2003; OFWAT, 2013) in several Western countries. The Italian Regulatory Authority for Energy, Networks and Environment, ARERA, has recently introduced a regulatory framework for measuring the technical quality of the integrated water service (ARERA, 2017). In particular, for the performance assessment of water supply infrastructures three indicators have been introduced, which are related to (1) water losses, (2) service interruptions and (3) water quality. The technical regulation is embodied in a penalty/reward competitive mechanism, which is based on rankings the utility performances, with direct impacts on their revenue and financial statements (Guerrini et al., 2020). The indicators related to water losses include the linear leakage index, $M1_a$, and the percentage leakage index, $M1_b$, drawing up jointly the macro-indicator $M1$ and its related "class", which is used by ARERA to assess the annual target reduction of water losses of each utility and to rank their performances.

Although AWWA and IWA rule out the use of the percentage leakage index (Alegre et al., 2016), it is still widely adopted, also to drive technical and management decisions. This work discusses in detail the reasons why the percentage leakage index $M1_b$ should not be used as WPI, being inappropriate for both benchmarking different systems and assessing the impacts of planning and maintenance activities, as well as driving the allocation of investments.

The main drawbacks of the percentage leakage index are two:

a) the mathematical formulation, because it includes the control variable (i.e. volume of water losses) at both numerator and denominator of the index ratio, which is as well affected by the stochasticity of water consumption at the denominator and its spatial distribution;

b) the definition of water consumption as part of total system inflow, whose ambiguity may result in dramatic changes in the index value, even resulting into physical unconsistencies in some common system layouts.

The thorough analysis based on such drawbacks demonstrates that the percentage leakage index is completely apart from the features of the physical system in hand, thus being not acceptable as the performance indicator to drive technical decisions. The discussion is supported by both the algebraic solution for a simplified WSS and the numerical analysis of a real WSS, carried out through advanced hydraulic modelling and mass-balance calibration, which allows a physically consistent representation of the leakage volumes. From such perspective, this work overcomes the limitations of drawing conclusions based on lumped statistics and/or qualitative reasoning only.

The conclusions unveil major negative impacts on management and planning of investments if driven by the percentage leakage index. They tackle a critical issue for water utilities and regulatory bodies, considering that this index has been adopted as WPI for targeting the investments funded by the REACT-EU post-pandemic plans (REACT-EU, 2021), the National Recovery and Resilience plans (PNRR) (Cerutti, 2021; Italiano, 2021), the Next Generation EU funding programme (Codogno & Van der Noord, 2022), other than those driven by the ARERA technical regulation.

## 2 WPIs for water losses: linear leakage index ($M1_a$) and percentage leakage index ($M1_b$)

From the hydraulic standpoint, real losses include the discharged volume from holes, cracks, and fittings of pipes, under pressure-dependent conditions. They include both leaks from reported and unreported bursts and undetectable outflows from fittings of mains and services (Lambert, 1994), namely "background leakages". Differently from reported bursts, background leakages and unreported bursts represent the main components of the total leakage volume since they often remain undetected long time before repair. Therefore, background leakages and unreported bursts are referred to also as *volumetric real losses* (Berardi et al., 2018) and the approaches used for the leakage outflow calculation rely on the Germanopoulos formulation (Germanopoulos, 1985) and FAVAD model (Van Zyl & Cassa, 2014), both assuming leakages as free orifices.

*Volumetric real losses* are, hence, *deterministic* pressure-dependent components of the real water losses, that do not cause abrupt changes in WDN hydraulic behaviour and may run for a long time with major volumetric effects on the global WDN mass balance (e.g. annual operating cycle). They provide a measure of the asset management quality because high *volumetric real losses* relate to asset deterioration and/or pressure excess over the values required for a correct and reliable service.

The linear leakage index and the percentage leakage index, namely $M1_a$ and $M1_b$ in the Italian regulation, are formulated as follows.

Considering a WSS, including water transport mains and distribution pipes, where $L_P$ [km] is the total length of the system pipes, and $W_{Leak}$ [m$^3$] is the total annual volume of water losses, the linear water losses indicator $M1_a$ [m$^3$/km/day] is defined as:

$$M1_a = \frac{1}{365} \cdot \frac{W_{Leak}}{L_P} \tag{1}$$

$W_{Leak}$ [m$^3$] can be calculated as the difference between the input annual water volumes, i. e. the total volumes entering the system, and the annual amount of the output volumes from the system, including authorized consumptions and uses, measured or unmeasured, billed or unbilled, and water delivered to other systems (ARERA, 2017).

In Eq. (1), $M1_a$ is defined as ratio between the main control variable ($W_{Leak}$) and the total extent of the system $L_p$. Since $L_p$ does not change over time, except for limited extensions/changes of the system, $M1_a$ represents a proxy of $W_{Leak}$ and thus can be effectively used to track the results of asset management, i.e. encompassing pipelines rehabilitation, pressure control or active leakage control activities. In addition, the reduction of $W_{Leak}$ achieved in one portion of the system, e.g. a DMA, linearly affects the reduction of $M1_a$ of the entire system.

The percentage water losses indicator $M1_b$ [%] is defined as ratio between $W_{Leak}$ [m$^3$] and the total input annual water volumes $W_{INP}$ [m$^3$]:

$$M1_b = \frac{W_{Leak}}{W_{INP}} = \frac{W_{Leak}}{W_{Leak} + D_p} \tag{2}$$

Eq. (2) explicitly shows that the total input volume $W_{INP}$ includes both the annual volume of water losses $W_{Leak}$ and the annual volume of the overall authorized water consumptions and uses $D_p$ [m$^3$]. The inconsistencies of $M1_b$, as mentioned in the introduction, are discussed in the following sections and demonstrated on a real case WSS.

### 2.1 Incosistency of $M1_b$ related to its formulation

The right-end side of Eq. (2) shows the management control variable $W_{Leak}$ at both numerator and denominator of the ratio. This circumstance makes the variation of $M1_b$ non linear with the variation of $W_{Leak}$, especially because in real systems $W_{Leak}$ and $D_p$ have the same order of magnitude.

In addition, $W_{Leak}$ represents a *deterministic* component of the water outflow, influenced by pressure and asset deterioration of the system, while $D_p$ encompasses the *stochastic* variability of water consumption, i.e. the fluctuations of water requests over time. This means that, even if the leakage volume $W_{Leak}$ does not change from one year to the next, $M1_b$ may increase or decrease according to the annual variation of water consumption $D_p$. Therefore, $M1_b$ is affected by socio-economic dynamics, as well as changes in consumption habits, as those observed during the recent pandemic restriction period (Spearing et al., 2021; Rohilla, 2020) or, in the long run, by policies to abate consumption waste in the framework of the sustainability goals for mitigating climate change effects. Besides, being $D_p$ at the denominator of Eq. (2) it results in $M1_b$ to be conflicting with such policies, since reducing water consumption leads to the increase (i.e. worsening) of the index.

Finally, from Eq.(2), $M1_b$ may be explicitly related to $M1_a$ as (Giustolisi & Mazzolani, 2022):

$$M1_b = \frac{W_{Leak}/L_P}{W_{Leak}/L_P + D_p/L_P} = \frac{M1_a}{M1_a + D1_a} \tag{3}$$

where $D1_a$ [m³/km/day] represents the density of water consumption, i.e. the daily volume of water consumption divided by the total length of the system pipes $L_p$:

$$D1_a = \frac{1}{365} \cdot \frac{D_p}{L_p} \tag{4}$$

Eq. (3) highlights the dependence of $M1_b$ to the stochasticity of $D1_a$, which is related to type of consumers, the variability of water requests over time and space, as well as the features of the system. Indeed, systems with low $D_p$ and/or large pipeline length $L_p$ (i.e. low $D1_a$) will show higher $M1_b$ and *viceversa*, irrespectively of $W_{Leak}$ and $M1_a$. From the asset management perspective, this means that higher $M1_b$ is based on annual water consumption $D_p$ and pipeline length $L_p$, which are features of the WSS that are not technical-decision variables in asset management actions.

Therefore, using $M1_b$ could be highly misleading to drive interventions and investments allocation of water utilities.

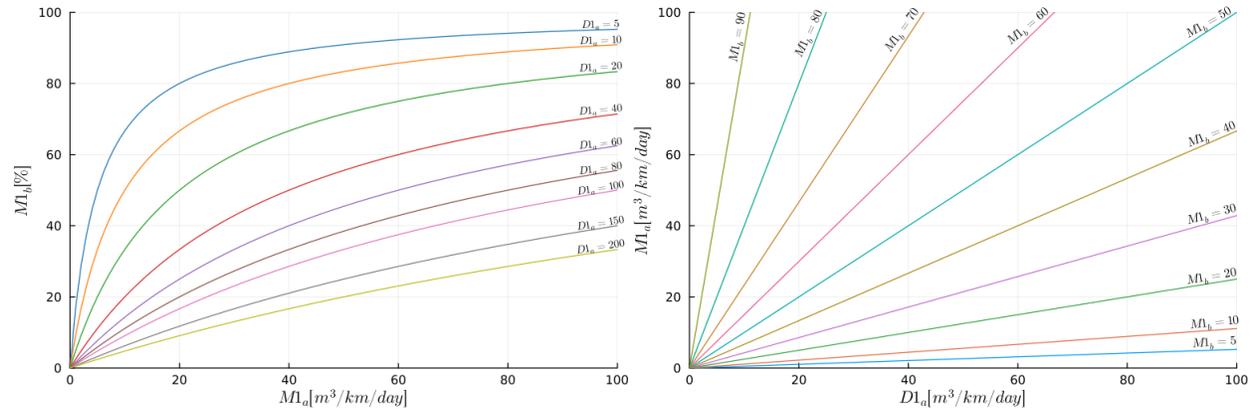

**Figure 1**. Representation of the $M1_b$ variability over $M1_a$ range, for assigned $D1_a$ [m³/km/day] (left) and representation of the $M1_a$ variability over $D1_a$ range, for assigned $M1_b$ [%] (right).

In Figure 1(left) Eq. (3) is plotted for different consumers density values $D1_a$, ranging from 5 to 200 m³/km/day, while Figure 1(right) shows the variability of $D1_a$ compared to $M1_a$, for a range of assigned $M1_b$ from 5 to 90 %.

Figure 1(left) shows the signficant influence of $D1_a$ on $M1_b$ maintaing a costant value of $M1_a$ (i.e. same volume of water losses for systems with the same length of pipes). For $M1_a = 20$ m³/km/day, $M1_b$ ranges from values above 50 % if $D1_a$ is lower than 20 m³/km/day to values below 20 % if $D1_a$ is higher than 80 m³/km/day. For low values of $D1_a$ and $M1_a$, a minimum increase of $M1_a$ (i.e. water loss volume) causes a large increase in $M1_b$. For low $M1_a$ values, ranging from 5 to 10 m³/km/day, $D1_a$ curves can be approximated to a linear relationship, thus halving $M1_a$ leads to halving $M1_b$. Conversely, for $M1_a > 10$ m³/km/day, $D1_a$ curves show a non-linear relationship, particularly for $D1_a$ below 100 m³/km/day. This means that, for low $D1_a$ and high $M1_a$, decreasing $M1_a$ does not lead to a proportional reduction of $M1_b$. For example, when halving $M1_a$ from 40 to 20 m³/km/day with $D1_a=10$ m³/km/day, $M1_b$ reduces only by 1/6, from 80% to 66.67%. Furthermore, high $M1_b$ values are associated to systems with low $D1_a$ (below 20-30 m³/km/day) even though the linear leakage index is quite satisfactory (10-20 m³/km/day), thus giving a flawed representation of the system state. On the other hand, for high $D1_a$ (above 80-100 m³/km/day) the relationship between $M1_a$ and $M1_b$ is almost proportional: indeed, when $D1_a = 150$ m³/km/day, halving $M1_a$ from 40 to 20 m³/km/day corresponds to a reduction of $M1_b$ of about 44 %, from 21.1 to 11.8 %. Besides, taking $D1_a = 150$ m³/km/day, an unsatisfactory level of linear leakage with $M1_a = 35$ m³/km/day corresponds to $M1_b = 18.9$ %, leading even in this case the flawed representation of a system with good leakage performances.

Such impacts are even more evident looking at Figure 1(right) where reducing $M1_b$ corresponds to moving from one line to the adjacent one. As $D1_a$ increases the distance between such lines, in terms of the difference between relevant $M1_a$, increases as well. For instance, moving from $M1_b = 50$ % to $M1_b = 40$ %, if $D1_a = 20$ m³/km/day corresponds to reducing $M1_a$ of 6.67 m³/km/day. Doubling $D1_a$ to 40 m³/km/day corresponds to the doubling of the $M1_a$ reduction as well, which turns to be equal to 13.34 m³/km/day.

Finally, the value of $D1_a$ also depends on the definition of $D_p$, which may drammatically change the value of $M1_b$, as discussed in the following section.

### 2.2 Incosistency of $M1_b$ related to the definition of system water cosumption

Eq. (3) shows that total system inflow has two components, namely the volume of water lost in the system $W_{Leak}$, which is the main control variable for asset management purposes, and the volume of water which is not lost and is designated as water "consumption" $D_p$. Nonetheless, defining the value of $D_p$ is not a trivial task since, depending on the adopted criterion, it may result into major changes in $M1_b$ values or even technical unconsistencies.

The ambiguity of defining $D_p$ depends on the accounting for the volumes which are transferred from the system to one or more other bordering systems, i.e. the water volume that passes through the system in hand without feeding any customers connected to it. In this case, the $M1_b$ indicator, based on the formulation in Eq. (3), can be written as:

$$M1_b = \frac{W_{Leak}}{W_{Leak} + D_p^{\{W\}} + D_p^{\{ext\}}} \qquad (5)$$

where $D_p^{\{W\}}$ represents the water requests by customers connected to the system, while $D_p^{\{ext\}}$ represents the volume crossing it and supplying *external* bordering systems. For instance, $D_p^{\{ext\}}$ is the volume passing through a transport pipeline to feed a WDN or the volume exiting from a WDN to feed another system downstream. It is worth remarking that $D_p^{[ext]}$ depends on the stochasticity of consumers demand and on leakages of the receiving system.

On one hand, if the definition of $D_p$ includes all the non-leakage volumes and it is assumed $D_p^{[ext]} \neq 0$, then the denominator of Eq. (5) will increase, thus reducing the value of $M1_b$. Since in a transport system $D_p^{[ext]}$ is by definition much larger than $D_p^{[W]}$ (being $D_p^{[W]} \approx 0$) its value may dramatically reduce $M1_b$. This is a technical inconsistency because $D_p^{[ext]}$ is independent of the leakage volume ($W_{Leak}$) of the system in hand (of which it lowers $M1_b$) but instead it is dependent on the leakage volume of the receving downstream system, in addition to its customer volume. Thus, the higher is the leakage volume of the downstream system, the smaller will be $M1_b$ of the system in hand, whose leakage representation is thus impaired. It is worth pointing out that similar inconsistencies are also met within water distribution networks if $D_p^{[ext]}$ is not much smaller then $D_p^{[W]}$. Considering two or more DMAs in a WDN, with comparable numbers of customers and leakage volumes (due to similar asset deterioration, pressure regime and DMA lengths), those that are downstream in the feeding scheme would result into higher values of $M1_b$ because they transfer lower or null volumes $D_p^{[ext]}$ downstream. As a consequence, investments on leakage management would be erroneously addressed towards such downstream DMAs since they would allow larger reduction of $M1_b$.

On the other hand, if the definition of $D_p$ includes only the customer volume then $M1_b$ is calculated with Eq. (5) without accounting for $D_p^{[ext]}$. Again, major technical inconsistency is met in transport systems, where $D_p^{[W]} \approx 0$ will result in $M1_b \approx 100$ %.

The above technical unconsistencies are not encountered using the linear leakage indicator $M1_a$, since no definition of $D_p$ is required in its formulation.

The ambiguity of defining $D_p$ also results into unconsistent calculation of $M1_b$ depending on the supplying scheme. Indeed, for the same DMA, $M1_b$ changes depending on whether it feeds a downstream DMA (i.e. series of DMAs, with $D_p^{[ext]} \neq 0$) or it is hydraulically disconnected from it (i.e. parallel DMA, with $D_p^{[ext]} = 0$). Such ambiguity further increases if DMAs are interconnected with each other allowing flow inversion in the connection pipes during normal operating cycle, which is a common occurrence in WDN operations.

Such circumstances may be observed with real WSSs, as reported in the following case study section.

## 3 Computing $M1_a$ and $M1_b$ in a real water supply system

The discussion on the two previously described WPIs is demonstrated on a real-life case study, represented by a water supply system feeding three contiguous small towns in Southern Italy, whose WDNs are named Net1, Net2, Net3.WDNs are fed by two main reservoirs, Res1 and Res2, through several supply pipelines, as shown in Figure 2.

For the purpose of the demonstration, a second possible configuration is considered assuming a pumping station from a well to be activated only for management purposes, in short periods of the year, at an inner node of the transmission pipelines. Such second configuration is also shown in Figure 2, where the reservoir Res3 represents the pressure provided by the pumping station supplying Net3, which is not connected to both Net1 and Net2.

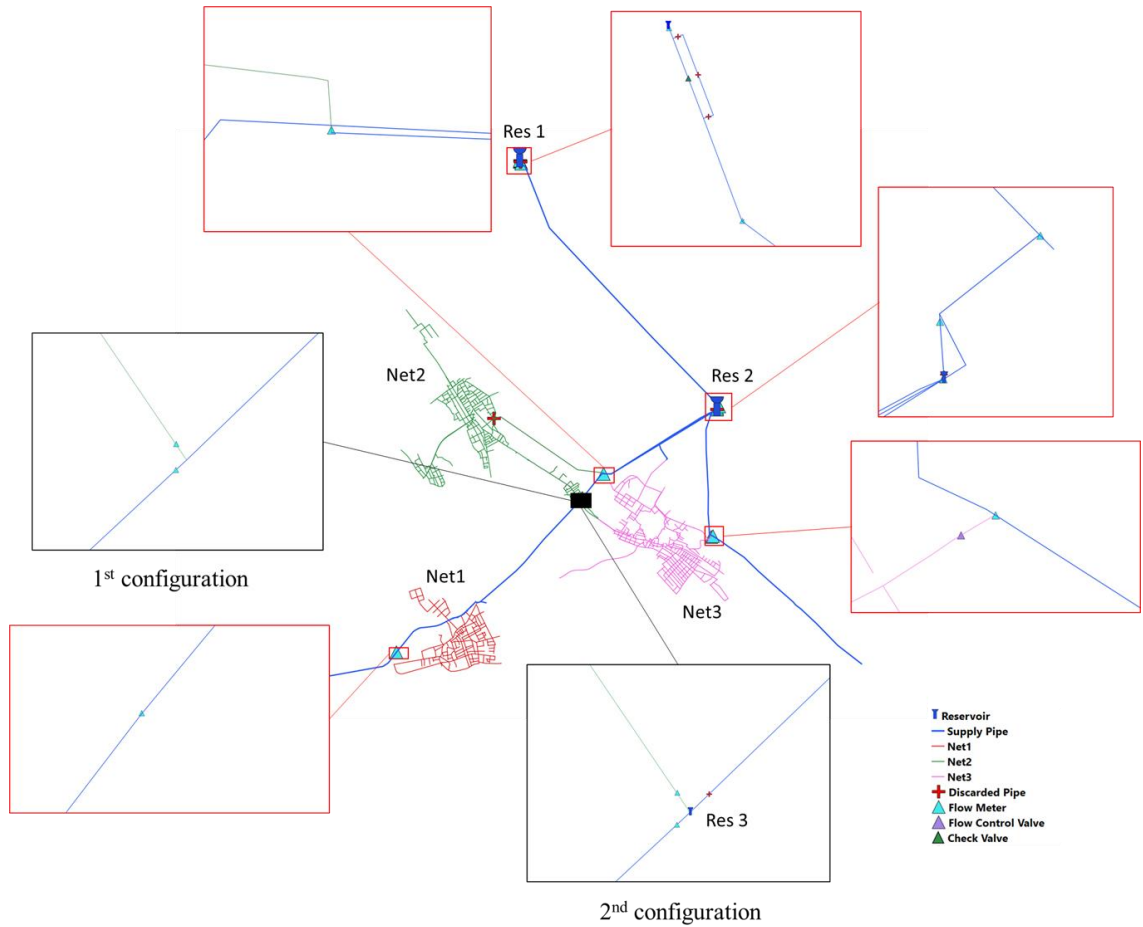

**Figure 2**. Representation of the layout for both case studies, comprising the transmission pipelines (blue lines), the three WDNs, Net1, Net2 and Net3, and the installed devices.

In both configurations, the main supply pipes are fed by a pump next to the Res1, which feeds Res2 and Net3 through a pressure reduction valve. A set of flow meters identifies the DMAs monitored close to inner nodes splitting the flow between WDNs, as will be discussed in Figure 5. Furthermore, the WSS delivers water out of the three WDNs; such external systems are identified in the hydraulic model as fictitious consumers located in the outer nodes, as shown in Figure 3. Each of them is associated with the water volumes feeding different tanks of the area, such as Ext1, Ext2, Ext3, Ext4, while Ext5 delivers water to a nearer WDN.

The water utility provided the hydraulic models of the three WDNs, including all the available information about topology, pipes asset and georeferenced position of water mains layout, customers and devices, taken from the utility GIS database, in addition to the reservoirs water level and flow rates records for the year 2021. For the same period the utility provided the annual customer consumptions as well as flow measurement and tank level records available as reported in Figure 2.

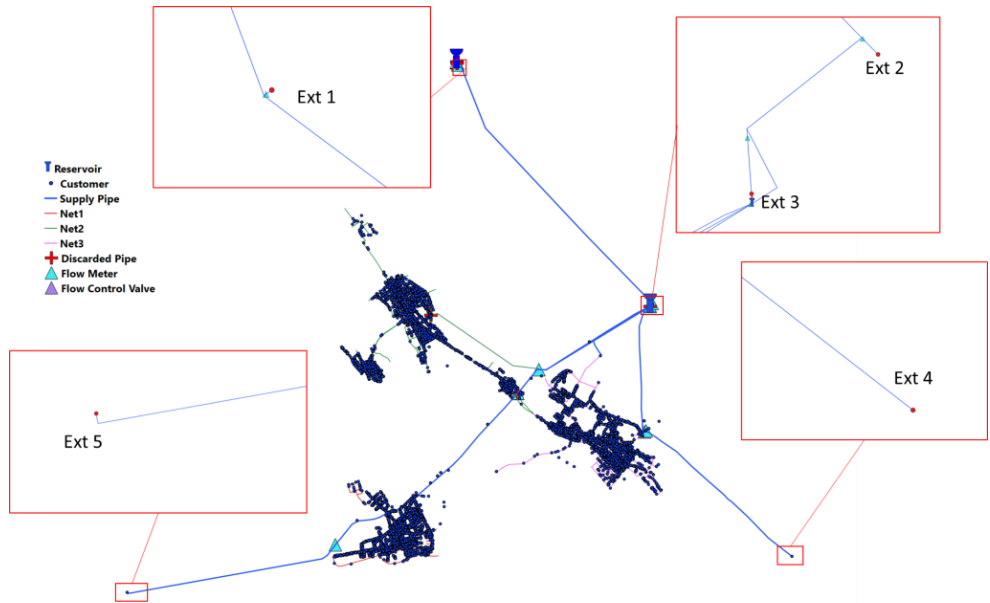

**Figure 3**. Representation of the case study layout, comprising devices and georeferenced positions of customers. The red dots indicate external outflows.

The total length of the system layout is about of 110 km, including 20 km of transport pipelines. The main information of the system layout, including length of each DMA and number of its customers, is shown in Table 1:

|            | L [km] | Customers [-] |
|------------|--------|---------------|
| Net1       | 23.46  | 4,125         |
| Net2       | 33.33  | 4,219         |
| Net3       | 34.92  | 4,902         |
| Water Mains| 18.84  | 27            |
| System     | 110.52 | 13,273        |

**Table 1.** Length of pipelines and number of customers for each part of the WSS case study.

The hydraulic analysis of the entire WSS has been performed using the WDNetXL-WDNetGIS software platform (WDNetXL-WDNetGIS, 2020). The core of the platform relies on the WDNetXL hydraulic solver, which allows a physically consistent representation of WDN hydraulics and leakages at single pipe level as a function of pressure and asset deterioration parameters (Giustolisi et al., 2008). In addition, as a relevant advancements to support asset management, the model calls data on single consumers from a separate database during its run, without need of preliminary aggregation of demands at nodes. The demand supplied to each customer is computed on the basis of pressure-dependent relationships which account for real type of connection, i.e direct connection, through private tanks, free orifices, at multi-storey buldings. The calibration of the hydraulic model has been performed using the same approach reported in Berardi & Giustolisi (2021), aimed at separating, from mass balance calculation at system or DMA level, pressure-dependent components of water outflows, i.e. consumer demands and leakages, based on inflow and pressure measurements. The methodology also allows identifying the calibrated demand pattern at both global network and DMAs level, and assessing the hydraulic resistances of most relevant pipes and parameters of the leakage model. For the sake of demonstration in this work, such calibration strategy was adopted to separate the leakage volume from customer volume and transferred to external supply.

The simulation uses flow and pressure measures recorded over the year 2021 by the water utility, with reference to five days (designated as OC in Table 2) representative of different characteristic operating states of the system: weekdays and summer holidays, weekdays and winter holiday and New Year's Day. Such approach allows overcoming the lack of recorded data in some days of the year, while providing a consistent representation of the system behaviour over one year. It also allows a more robust modelling than using a limited number of field data measured in specific days and locations, usually not synchronous with the water consumption time series being referred to the previous year.

| Operating Cycle | Inflow [m³] | Demand [m³] | Leakage [m³] | $M1_a$ [m³/km/d] | $M1_b$ [%] | $D1_a$ [m³/km/d] |
|---|---|---|---|---|---|---|
| OC 1 | 6,539 | 2,910 | 3,628 | 32.82 | 55.49% | 26.32 |
| OC 2 | 6,403 | 2,615 | 3,788 | 34.26 | 59.16% | 23.65 |
| OC 3 | 5,892 | 1,673 | 4,219 | 38.16 | 71.61% | 15.13 |
| OC 4 | 5,870 | 1,604 | 4,266 | 38.58 | 72.68% | 14.50 |
| OC 5 | 6,208 | 2,061 | 4,147 | 37.50 | 66.80% | 18.64 |
| Mean OC | 6,182 | 2,173 | 4,010 | 36.26 | 64.86% | 19.65 |

**Table 2.** Results of the water balance terms of the system, i.e. inflow, demand and leakage, and the calculation of WPIs for each operating cycle.

Table 2 reports the water balance figures and the calculation of $M1_a$ and $M1_b$ of the system, with reference to the different operating cycles (OCs) and their average values. It is worth noting that the advanced modelling and calibration approach allows identifying the variation of the leakage volume as a result of the change of pressure regime due to different daily water consumptions.

Even considering $M1_b$ at daily scale, it is strongly related to the variability of $D1_a$, which turns into a much higher variation of the indicator compared to $M1_a$. Conversely, the variation of $M1_a$ is independent of $D1_a$ being dependent linearly on the leakage volume, which in turn increases following the slightly higher pressure level due to lower demand (and viceversa). In more details, considering OC4 and OC5, for an increase of about 4 m³/km/day of $D1_a$ (i.e. 21% of the mean value of $D1_a$ of the five OCs) it turns into the unfair reduction of about 8,1% of the $M1_b$ value (from 72.68 to 66.80%), compared to 2.8% reduction of $M1_a$ (from 38.58 to 37.50 m³/km/d), which instead reflects the same percentage decrease of the leakage volume (from 4,266 to 4,147 m³/d). Considering that OC4 and OC5 represents two operating cycle of the same season, it is remarked that fluctuations of the customer demands density confirms higher fluctuations of $M1_b$ than $M1_a$. This result gives evidence that $M1_a$ is a more representative indicator for defining the leakage level in the system, even at lower time scales.

If the same analysis is carried out comparing the operating cycles during summer holidays (OC1) and winter (OC4), $M1_b$ rises up to 31% due to the consumer demand decrease, while $M1_a$ increases of about 17,6 % as a consequence of the seasonal change of the pressure regime, which in turn leads to an increase of the leakage volume of the same percentage.

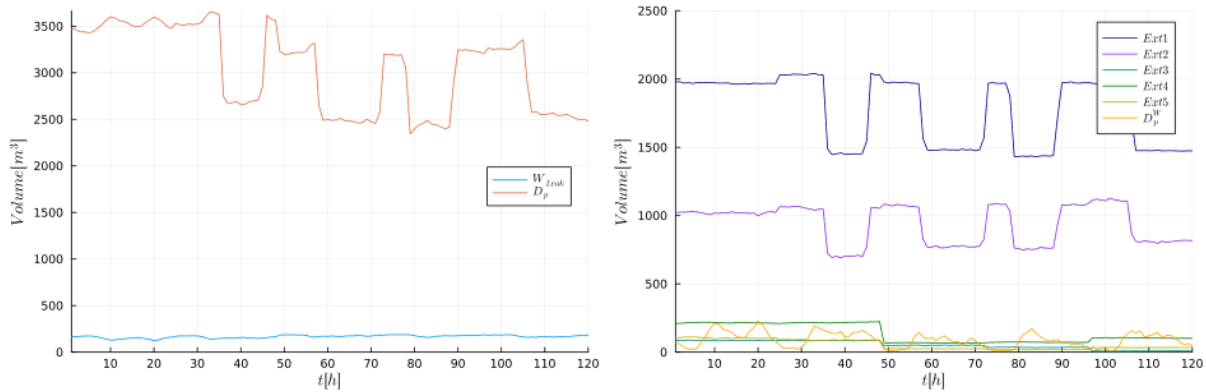

**Figure 4**. Plots of the outlet volumes evaluated after the hydraulic simulation (left), and of the variability of the customers demands of WSS and the external consumers nodes. (right).

Figure 4 shows two different plots of the simulated outlet volumes of the case study, with indication of leakage and consumption volumes over the five OCs, on the left, and the variability of the external demands of the WSS, on the right. It can be shown that the highest requests of water by the external consumers relies on Ext1 and Ext2, which are located at the outer nodes of the transmission pipelines between Res1 and Res2. The hydraulic simulation of WSS for both configurations shown in Figure 2 and the related outlet volumes are similar, even if the supplying schemes are different, with a minimum decrease of the leakage volume in Net2 and Net3, due to slight pressure increase in the second configuration.

$M1_a$ and $M1_b$ have been computed also for each DMA identified by existing flow meters. Figure 5 shows DMAs of the system, where DMA#1 comprises Net2 and Net3 and the surrounding transmission mains, while Net1 is delimited by its own district, DMA#1. DMA#3 and DMA#4 are "fictious" districts, related to the outer mains without customers.

The hydraulic model was used to perform the advanced hydraulic analysis of the whole system and its DMAs. The effective DMAs for the evaluation of the hydraulic behavior of the system are represented by DMA#1 and DMA#2 for both supplying configurations, i.e. in normal operating condition (without Res3) and assuming pumping (with Res3), neglecting DMA#3 and DMA#4 whose aggregated outflows have beed accounted to perform the consistent hydraulic analysis of the whole system.

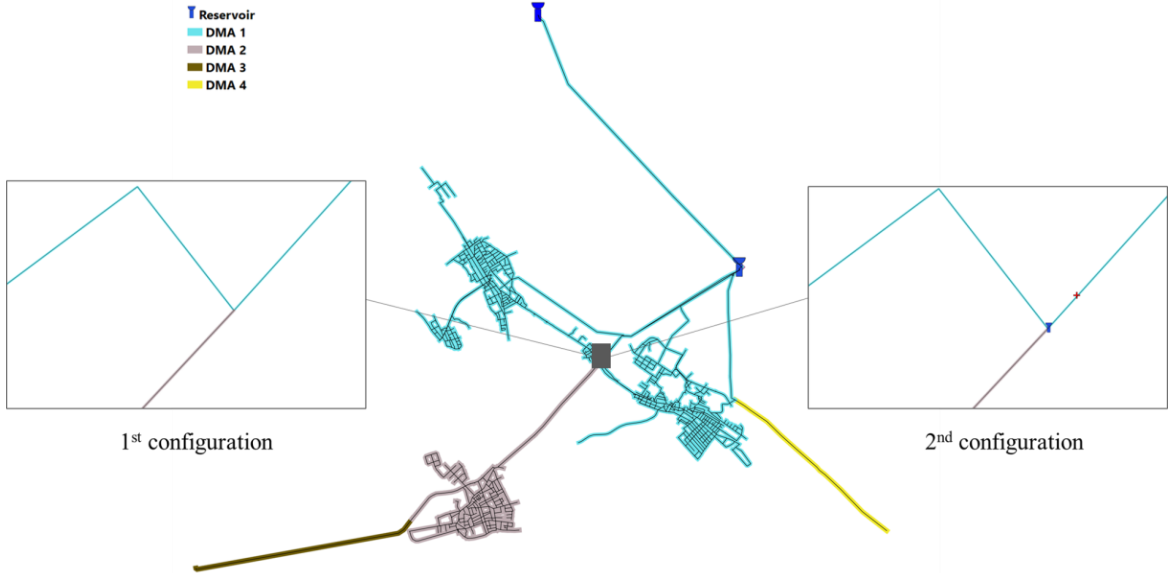

**Figure 5**. Representation of the DMAs of the system for both supplying configurations.

The main hydraulic parameters of the first supplying configuration of Figure 2, including the annual leakage and consumption volumes as expressed in Eq.(5), are shown in the Table 3, where the value of $M1_b$ has been calculated assuming the two possible definitions of $D_p$, i.e. assuming $D_p^{[ext]} = 0$ or $D_p^{[ext]} \neq 0$ in Eq. (5).

|  | DMA #1 | DMA #2 | System |
|---|---|---|---|
| Length [km] | 84.76 | 25.79 | 110.55 |
| Average Pressure [m] | 49.2 | 65.4 | 53.1 |
| $W_{Leak}$ [m³] | 1,052,144 | 411,393 | 1,463,527 |
| $D_p^{[W]}$ [m³] | 519,618 | 273,410 | 793,028 |
| $D_p^{[ext]}$ [m³] | 26,512,821 | 474,531 | 25,828,017 |
| $M1_a$ [m³/km/d] | 34.01 | 43.71 | 36.27 |
| $D1_a$ [m³/km/d] ($D_p^{[ext]}$=0) | 16.79 | 29.04 | 19.65 |
| $M1_b$ [%] ($D_p^{[ext]}$=0) | 66.98 | 60.08 | 64.86 |
| $D1_a$ [m³/km/d] ($D_p^{[ext]} \neq 0$) | 873.80 | 79.47 | 659.78 |
| $M1_b$ [%] ($D_p^{[ext]} \neq 0$) | 3.75 | 35.48 | 5.21 |

**Table 3.** Results of WSS and DMAs hydraulic parameters of the first supplying configuration.

In the case of $D_p^{[ext]} = 0$, the consumption is only the water supplied to customers connected to each DMA. DMA#2 can be viewed as a WDN with a lower customer density value than DMA1, and both exhibit a greater variability of $M1_b$ compared to the $M1_a$. Such variability can affect the technical and economical priorities of leakage reduction of the system at DMA level. The $M1_b$ indicator of the system is greater than the $M1_b$ values of each DMA, confirming that the percentage indicator is not a consistent indicator for leakage assessment of the whole WSS.

In case of $D_p^{[ext]} \neq 0$, the water consumption is meant as the total volume entering the system or the DMA, except for leakage. In this case $M1_b$ of the WSS rapidly decreases, with an increase of $D1_a$, due to the higher amount of water volumes delivered to the outer systems. It is worth to remark that $D_p^{[ext]}$ value also depends on leakage and customer volumes of DMA#3 and DMA#4.

The same conclusions can be drawn for the two DMAs when changing the assumption on $D_p^{[ext]}$: $M1_b$ in DMA#1 decreases from 66.98% to 3.75%, while $M1_b$ in DMA#2 decreases from 60.08% to 35.48%.

Table 4 shows the same data as Table 3 related to the second supplying configuration of Figure 2. Table 3 and Table 4 show that $M1_a$ value at system and DMAs level remains the same for both cases, because it depends only on the amount of water losses from each pipe, thus highlighting the spatial dependence of the indicator.

In the second supplying scheme, the hydraulic behavior of each DMA is similar to the previous case, but the supplying layout has changed because DMA#1 is fed by Res1, Res2 and Res3, while DMA#2 by the Res3 only.

Since the water balance is the almost same for both configuration, $M1_a$ and $M1_b$ indicators at system level remain the same, as from comparing Table 3 and Table 4, where the only difference stands from the small increase in $W_{Leak}$ in DMA#1.

At DMAs level $M1_a$ remains unchanged, being dependent only on the leakage volume, while $M1_b$ for DMA#1 increase from 3.75 to 3.91 % (for $D_p^{[ext]} \neq 0$), because DMA#1 no longer feeds DMA#2. This happens because the volume crossing DMA#1, i.e. feeding DMA#2, in the previous case is no longer computed in the $M1_b$ of DMA#1, and the water volume is supplied by Res3 to DMA#2. It is worth nothing that, for $D_p^{[ext]} = 0$, $M1_b$ remains the same for both configurations, since it depends on customer consumptions only.

| | DMA #1 | DMA #2 | System |
|---|---|---|---|
| Length [km] | 84.76 | 25.79 | 110.55 |
| Average Pressure [m] | 49.5 | 65.4 | 53.1 |
| $W_{Leak}$ [m$^3$] | 1,054,000 | 411,393 | 1,463,527 |
| $D_p^{[W]}$ [m$^3$] | 519,618 | 273,410 | 793,028 |
| $D_p^{[ext]}$ [m$^3$] | 25,353,485 | 474,531 | 25,828,017 |
| $M1_a$ [m$^3$/km/d] | 34.07 | 43.71 | 36.27 |
| $D1_a$ [m$^3$/km/d] ($D_p^{[ext]}$=0) | 16.79 | 29.04 | 19.65 |
| $M1_b$ [%] ($D_p^{[ext]}$=0) | 66.98 | 60.08 | 64.86 |
| $D1_a$ [m$^3$/km/d] ($D_p^{[ext]}\neq0$) | 836.33 | 79.47 | 659.78 |
| $M1_b$ [%] ($D_p^{[ext]}\neq0$) | 3.91 | 35.48 | 5.21 |

**Table 4.** Results of WSS and DMAs hydraulic parameters of the second supplying configuration.

## 4 Algebraic equations of M1ₐ and M1_b in transport and distribution system layouts

Figure 6a shows a system layout consisting of a water transport system without water connections and a WDN, which are schematized as a water main fed by a reservoir carrying water to the WDN tank. The output volume from WDN, $W^{[D]}_{OUT}$, is equal to the sum of customers demands and other authorized volumes. Without impairing the generality of the analysis, it is assumed no overflow from the WDN tank, thus $W_{OUT}^{[T]} = W_{INP}^{[D]}$. In such system, for which we calculate the two leakage indicators, $M1_a$ and $M1_b$, input and output volumes of transport pipe and WDN are known, and by difference water losses $W_{Leak}^{[T]}$ and $W_{Leak}^{[D]}$, assumed to be greater than zero in both schemes.

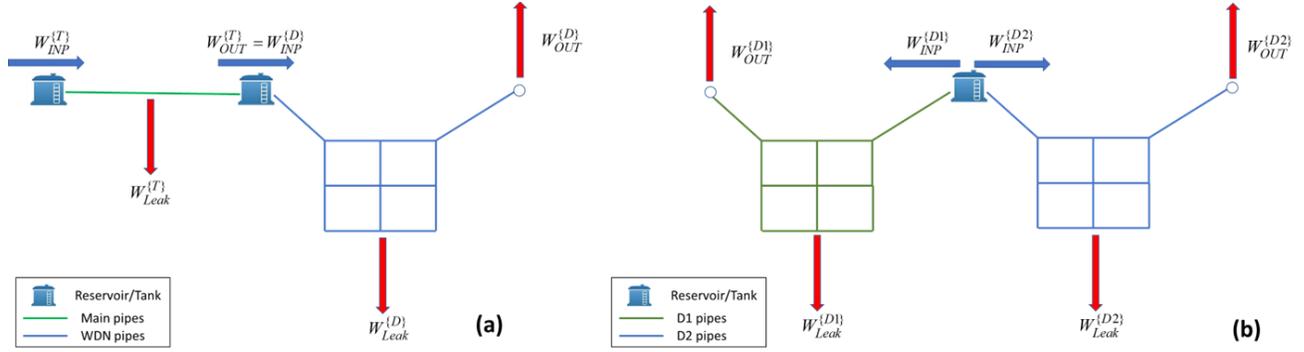

**Figure 6**. Two different schematic system layouts: volumes feeding the two systems (blue arrows); volume leaving the system, i.e. authorized consumption and leakages (red arrows).

In this system, $M1_a$ indicator for both transport main and WDN are:

$$M1_a^{\{T\}} = \frac{W_{Leak}^{\{T\}}}{L_p^{\{T\}}} = \frac{W_{INP}^{\{T\}} - W_{INP}^{\{D\}}}{L_p^{\{T\}}} \tag{6}$$

$$M1_a^{\{D\}} = \frac{W_{Leak}^{\{D\}}}{L_p^{\{D\}}} = \frac{W_{INP}^{\{D\}} - W_{OUT}^{\{D\}}}{L_p^{\{D\}}} \tag{7}$$

$M1_a$ of the unified transport and distribution system, $M1_a^{\{T \cup D\}}$, turns to be:

$$M1_a^{\{T \cup D\}} = \frac{W_{Leak}^{\{T\}} + W_{Leak}^{\{D\}}}{L_p^{\{T \cup D\}}} = \frac{L_p^{\{T\}} \cdot M1_a^{\{T\}} + L_p^{\{D\}} \cdot M1_a^{\{D\}}}{L_p^{\{T \cup D\}}} \tag{8}$$

where $L_p^{\{T \cup D\}} = L_p^{\{T\}} + L_p^{\{D\}}$. Thus, the linear indicator $M1_a$ of the whole system, $M1_a^{\{T \cup D\}}$, is equal to the average of $M1_a$ of each part of the system weighted by the relevant pipeline lengths. Thus, $M1_a^{\{T \cup D\}}$ is more influenced by the longer part of the water supply system, usually the WDN.

On the other hand, $M1_b$ of the unified transport and distribution system, $M1_b^{\{T \cup D\}}$, is equal to:

$$M1_b^{\{T \cup D\}} = \frac{W_{Leak}^{\{T \cup D\}}}{W_{INP}^{\{T \cup D\}}} = \frac{\left(W_{INP}^{\{T\}} - W_{OUT}^{\{T\}}\right) + \left(W_{INP}^{\{D\}} - W_{OUT}^{\{D\}}\right)}{W_{INP}^{\{T\}}} \tag{9}$$

Being $W_{OUT}^{\{T\}} = W_{INP}^{\{D\}}$, and the total input volume $W_{INP}^{\{T \cup D\}} = W_{INP}^{\{T\}}$, $M1_b^{\{T \cup D\}}$ turns to:

$$M1_b^{\{T \cup D\}} = \frac{W_{Leak}^{\{T\}} + W_{Leak}^{\{D\}}}{W_{INP}^{\{T\}}} = \frac{W_{INP}^{\{T\}} - W_{OUT}^{\{D\}}}{W_{INP}^{\{T\}}} \tag{10}$$

Thus, the percentage leakage index of the whole system comprising transport system and WDN is equal to the ratio of the difference between the input volume in the transport system and the output volumes from the WDN divided by the input volume in the transport system. Eq. (10) can be rearranged to express the $M1_b$ indicator of the whole system as a function of $M1_b^{\{T\}}$ and $M1_b^{\{D\}}$:

$$MI_b^{\{T \cup D\}} = MI_b^{\{T\}} + MI_b^{\{D\}} \frac{W_{OUT}^{\{T\}}}{W_{INP}^{\{T\}}} > MI_b^{\{T\}} \tag{11}$$

$$MI_b^{\{T \cup D\}} = MI_b^{\{D\}} + MI_b^{\{T\}} \frac{W_{OUT}^{\{D\}}}{W_{INP}^{\{D\}}} > MI_b^{\{D\}} \tag{12}$$

Under the assumption that water losses of the two systems are greater than zero, the ratios in (11) and (12) are always positive and less than 1. Therefore, $MI_b$ indicator of the whole system is always greater than both $MI_b$ indicators of each part of the system. Hence:

$$MI_b^{\{T \cup D\}} > \max\left(MI_b^{\{T\}}, MI_b^{\{D\}}\right) \tag{13}$$

It can be shown that $MI_b^{\{T \cup D\}}$ is also less than the sum of $MI_b^{\{T\}}$ and $MI_b^{\{D\}}$.

The result achieved by (13) underlines that $MI_b^{\{T \cup D\}}$ fails to be a consistent indicator because the value of the whole system is not in-between those of the single transport and distribution systems, being always larger than both. Thus, $MI_b^{\{T \cup D\}}$ cannot be regarded as a scalable indicator, meaning that it does not allow the comparison of different water supply systems, being undermined by their spatial extension and hydraulic connections, as also shown in the case study analysis.

Figure 6b shows, using the same symbols of Figure 6a, a different supplying layout composed of two WDNs, D1 and D2, fed separately by a single tank. In this configuration, $MI_b$ indicator of the whole system, $MI_b^{\{D1 \cup D2\}}$, is:

$$MI_b^{\{D1 \cup D2\}} = \frac{W_{Leak}^{\{D1\}} + W_{Leak}^{\{D2\}}}{W_{INP}^{\{D1\}} + W_{INP}^{\{D2\}}} \tag{14}$$

That, in terms of $MI_b$ of each WDN is written:

$$MI_b^{\{D1 \cup D2\}} = \frac{MI_b^{\{D1\}} W_{INP}^{\{D1\}} + MI_b^{\{D2\}} W_{INP}^{\{D2\}}}{W_{INP}^{\{D1\}} + W_{INP}^{\{D2\}}} \tag{15}$$

Unlilke the previus configuration of transport and distribution schemes connected each other as a "series" system, in this configuration the two schemes may be regarded as a "parallel" system, whose unified $MI_b^{\{D1 \cup D2\}}$ is not undermined by the system extenson being equal to the average of the $MI_b$ of each WDN weighted on the input volumes of each system.

On the other hand, the unified linear indicator $MI_d^{\{D1 \cup D2\}}$ of this parallel system configuration maintains the same weighted average found for the series system configuration (Eq. 8), because it is independent on the water systems layout, being dependent only on the leakage volume of each system and their lengths.

## 5 Conclusions

WPIs are used for quantifying the status of water supply infrastructures in terms of management and operative actions, and thus can be used as leakage performance parameters for both targeting the efforts that water utilities need to face and benchmarking their results. The recent

REACT-EU and Next Generation EU funding programmes have triggered investments for the ecological transition and infrastructures sustainability, applying the percentage index $M1_b$ as performance indicator. $M1_b$ is as well applied by the Italian Regulatory Authority (ARERA) along with the linear leakage index $M1_a$ in regulating the technical quality of the water service. The values of both indexes are thus used to establish the annual target of the $M1_a$ reduction of each water utility, whose leakage performances are then ranked and evaluated in a penalty-reward mechanism, along with other WPIs.

As discussed in this paper, the leakage percentage indicator reveals several drawbacks, summarized as follows.

- The formulation of $M1_b$ shows the same control variable, i.e. the volume of leakages along the system, both at numerator and denominator, resulting in a non-linear relationship between $M1_b$ and the leakage volume.

- $M1_b$ is strictly dependent on the water consumption density, which represents itself a variable of the water balance severely affected by the demand fluctuations over time. Its formulation results into the technical nonsense of increasing $M1_b$ by decreasing the water consumption. Thus, water utilities are discouraged to promote actions for reducing consumption waste and avoid improper use, for instance sharing with customers the alarm of user-side leakages, which is today a common functionality of any electronic water meter.

- Due to its dependency on customer consumption, $M1_b$ severely overestimates leakage level for systems or network with low consumptions density and underestimates it for high consumptions density, which turns to undermine the priorities of investments for asset management.

- The definition of water consumption at denominator of $M1_b$ is ambiguous in itself. If the water volume delivered from a system/DMA to a bordering one is included in the water consumption density, $M1_b$ turns to be lowered. Conversely, when such volume is not included in the formulation, $D_p$ decreases and $M1_b$ increases. In the case transport systems, which should be taken into account in the leakage management activities along with WDNs, the $M1_b$ calculation turns to be about 100 %, which is a technical nonsense.

- $M1_b$ is strictly dependent on system supplying configuration. Assuming the case study with two different supplying schemes, one with DMAs fed in cascade and the other with DMAs fed independently, it has been reported that $M1_b$ will increase for the upper DMAs of the former scheme, even though the $M1_b$ values at the entire system level remain the same for both schemes. Thus, applying $M1_b$ at the system level may be misleading with respect to asset management activities, especially for utilities with long transport system and/or WDN fed in cascade.

- The algebraic calculation of $M1_b$ for the unified system of transport and distribution schemes shows the inconsistency of the indicator, since its value turns to be greater than both the indexes of the single schemes, instead of being in-between, as expected by a scalar WPI.

Such features make the percentage leakage index $M1_b$ unsuitable for both assessing the leakage management performances and benchmarking or ranking different systems/DMAs. Therefore, it should not be utilized in order to avoid the risk of ineffective actions and inefficient, not to say improper, allocation of financial resources for investments.

Conversely, it has been confirmed that the linear leakage indicator $M1_a$ is consistent with the physically-based behavior of the water systems, since its value is dependent neither on the customer density nor on the system supplying layout. These features make $M1_a$ suitable for benchmarking leakage management performances and assessing the expected impacts of asset management and planning tasks.

**Acknowledgments**